\documentclass[preprint,showpacs,preprintnumbers,amsmath,amssymb,tightenlines,nofootinbib]{revtex4-1}
 \usepackage{graphicx}
 \usepackage{dcolumn}
 \usepackage{bm}
 \usepackage{ulem}
 \usepackage{enumitem}
 \usepackage{slashed}
 \usepackage{amssymb}

\def\p{\bm{p}}
\def\k{\bm{k}}
\def\v{\bm{v}}
\def\x{\bm{x}}

\def\b{\bm{b}}

\def\vec\epsilon{\bm{\epsilon}}

\def\p{\bm{p}}
\def\k{\bm{k}}
\def\v{\bm{v}}
\def\x{\bm{x}}

\def\u{\bm{u}}

\usepackage{xcolor}

\begin{document}

\title{Heavy quark suppression and anisotropic flow at intermediate momentum}
\author{Juhee Hong}
\affiliation{Department of Physics and Institute of Physics and Applied Physics, Yonsei University,
Seoul 03722, Korea}
\date{\today}

\begin{abstract}
In an intermediate-momentum regime where mass effects are  
significant, heavy quark suppression and anisotropic flow are computed to 
investigate the transition between the collisional and radiative energy 
loss.   
Based on the collision kernel for diffusion, 
elastic scattering and semi-collinear gluon-bremsstrahlung can be 
consistently incorporated into a Boltzmann equation that involves the heavy 
quark diffusion coefficient. 
Using the running coupling constant and the diffusion coefficient 
constrained by lattice QCD data, the collisional and radiative energy-loss 
contributions to the $R_{AA}$ and $v_2$ are studied in 
hydrodynamically expanding thermal media. 
The evolution of the observables, the bulk flow effect, and the 
dependence on mass and centrality are discussed in noncentral heavy-ion 
collisions. 
\end{abstract}

\maketitle

\section{Introduction}

Heavy quark energy loss and jet quenching are important phenomena to 
understand the transport properties of quark-gluon plasmas(QGP)  
in relativistic heavy-ion collisions. 
Using various formalisms of energy loss, several transport models of heavy 
quarks have been developed, most of which, with some adjustment to model 
parameters, reasonably describe experimental data of heavy mesons
\cite{Andronic:2015wma,Dong:2019byy,Apolinario:2022vzg,He:2022ywp}. 
Some of these models consider the collisional energy loss by elastic scattering 
only, while others also include the radiative one by gluon-bremsstrahlung. 
It is generally understood that heavy quark energy loss is dominated by 
elastic scattering and gluon emission at low and high momenta, respectively 
\cite{Andronic:2015wma,Dong:2019byy,Apolinario:2022vzg,He:2022ywp}. 
However, the intermediate-momentum region, where the heavy quark mass is 
nonnegligible, has large uncertainties regarding the significance of the 
radiative effect and the momentum-dependent transition between the relative 
dominance of the two mechanisms. 
Heavy quarks with finite mass and moderate momentum are particularly unique 
probes for studying the different energy-loss effects in QGP. 
To investigate the heavy quark energy loss in the intermediate-momentum region, 
a transport approach has recently been developed based on a Boltzmann equation 
with elastic scattering and semi-collinear gluon emission \cite{Hong:2023cwl}. 
This approach enables distinct treatments of diffusion and radiation, 
while still describing both processes consistently 
through a single transport parameter, the heavy quark diffusion coefficient.

Heavy flavor interactions in strongly interacting QCD matter have mainly been 
studied with two experimental observables, the nuclear modification 
factor($R_{AA}$) and the elliptic flow($v_2$) which quantify medium 
modifications and momentum-space anisotropy, respectively.  
These observables are affected by numerous factors which include initial heavy 
quark production, expansion of background media, heavy quark 
interaction with QGP, hadronization, and hadronic rescattering. 
As a primary contribution to heavy flavor suppression, 
the heavy quark $R_{AA}$ by energy loss has been investigated in 
Ref. \cite{Hong:2023cwl}. 
This paper complements the previous work by applying the same formalism 
to noncentral heavy-ion collisions in order to compute both the suppression 
factor and the anisotropic flow of heavy quarks, which are induced by two 
types of energy loss. 
Although various approaches have been developed to estimate the degree of heavy 
quark energy loss and simultaneously describe $R_{AA}$ and $v_2$ of 
heavy flavor, heavy quark modeling remains challenging 
\cite{Djordjevic:2013pba,Das:2015ana}. 
While the suppression factor is well-reproduced by transport models, the 
elliptic flow may be underestimated \cite{Andronic:2015wma}. 
Compared to the $R_{AA}$, the $v_2$ is more complicated by initial 
fluctuations, bulk flow, and freezeout. 
Since there are multiple sources of azimuthal anisotropies through all stages 
of heavy-ion collisions, it is crucial to understand the energy-loss 
contribution to the heavy quark flow in the QGP phase.

The goal of this paper is to qualitatively examine how different types of 
energy loss contribute to the heavy flavor observables, $R_{AA}$ and $v_2$, 
in an intermediate-momentum regime. 
Sec. \ref{formalism} begins with a brief review of heavy quark transport 
formulated in Ref. \cite{Hong:2023cwl}. 
In Sec. \ref{result}, the nuclear modification factor and the elliptic flow of 
heavy quarks are calculated in noncentral heavy-ion collisions simulated with 
relativistic hydrodynamics. 
In expanding QGP, the evolution of two observables, the collisional and 
radiative energy-loss contributions, the bulk flow effect, and the 
dependence on impact parameter and mass are analyzed using the heavy quark 
diffusion coefficient which is constrained by lattice QCD data.  
Finally, Sec. \ref{summary} provides a summary.

\section{Heavy quark diffusion and radiation}
\label{formalism}

Heavy quarks undergo Brownian motion at low momentum, while 
medium-induced gluon emission dominates in the high-momentum limit.  
To study the transition from diffusion to radiation at intermediate 
momentum, a transport approach involving the heavy quark 
diffusion coefficient has been introduced in Ref. \cite{Hong:2023cwl}. 
A brief review and the final form of the transport equation are provided 
in this section.

The heavy quark distribution function is determined by a Boltzmann equation 
with the collisional and radiative energy-loss terms, 
\begin{equation}
\label{BoltzEq}
\left(\frac{\partial}{\partial t}+\v\cdot\frac{\partial}{\partial\x}\right)
f(t,\x,\p)
=C_{\rm col}[f]+C_{\rm rad}[f] \, .
\end{equation}
For soft momentum transfer, the collisional term can be approximated by a 
Fokker-Planck operator \cite{Svetitsky:1987gq}
\begin{equation}
C_{\rm col}[f]=\frac{\partial}{\partial p^i}\left[\eta(\p)p^if(\p)\right]
+\frac{1}{2}\frac{\partial^2}{\partial p^i\partial p^j}
\left[\kappa^{ij}(\p)f(\p)\right] \, ,
\end{equation}
where $\eta(\p)$ is the drag coefficient and 
$\kappa^{ij}(\p)=\kappa_L(p)\hat{p}^i\hat{p}^{j}+\kappa_T(p)(\delta^{ij}-\hat{p}^i\hat{p}^{j})$ is the momentum diffusion tensor. 
The drag coefficient is related to the longitudinal diffusion coefficient, 
$\eta(p)=\kappa_L(p)/(2TE_{\p})+\mathcal{O}(T/E_{\p})$, by the requirement of 
thermodynamic equilibrium in the large-time limit.

The formation time of gluon is given by the energy change in 
gluon-bremsstrahlung, $t_f= 2k(1-x)/(\k_T^2+m^2x^2+m_g^2)$, where 
$m$ is the heavy quark mass, $x=k/E_{\p+\k}$, $m_g^2=m_D^2/2$ is the thermal 
mass of gluon, $E_{\p+\k}$ is heavy quark energy, $k\sim T$ and $k_T$ are 
gluon energy and transverse momentum, respectively. 
To smoothly interpolate between low- and high-momentum regimes, a kinematic 
region where $gT\ll mx\ll T$ is considered. 
Since the formation time is shorter than the mean free path in this region, 
the gluon emission from a single scattering is 
used to compute the radiative term. 
In the semi-collinear limit, the gluon emission rate is  
\cite{Arnold:2002ja,Ghiglieri:2013gia,Ghiglieri:2015zma,Ghiglieri:2015ala}
\begin{multline}
\frac{d\Gamma(E_{\p},k)}{dk}=
\frac{g^2C_F\kappa_{T}(p)}{\pi k}[1+n_B(K)][1-n_F(P-K)]
\\
\times [(1-x)^2+1]
\int \frac{d^2\p_T}{(2\pi)^2}\frac{1}{(\p_T^2+m^2x^2+m_g^2)^2} \, ,
\end{multline}
where $\Gamma(E_{\p},k)$ is the rate for a heavy quark with momentum $\p$ to 
emit a gluon with energy $k$, and 
$n_{B,F}(K)=[\exp(K\cdot U/T)\mp 1]^{-1}$ (with four-momentum 
$K=(E_{\k},\k)$ and fluid velocity $U=(\gamma,\gamma\u)$, where 
$\gamma=1/\sqrt{1-\u^2}$ is the Lorentz factor) which corresponds to the 
Bose-Einstein or Fermi-Dirac thermal distribution in the plasma rest frame 
(for heavy quark, $n_F(P-K)$ is negligible). 
The emission rate has been found to be proportional to the diffusion 
coefficient, assuming that gluon has larger transverse 
momentum than soft momentum transfer. 
If $m^2x^2$ is greater than $\p_T^2$, 
gluon emission is suppressed at smaller angle than $\sim m/E_{\p}$ 
\cite{Dokshitzer:2001zm}. 
In Eq. (\ref{BoltzEq}), the radiative energy loss is taken into account as 
follows: 
\begin{equation}
C_{\rm rad}[f]= 
\int dk \left[
f(\p+\k)\frac{d\Gamma(E_{\p+\k},k)}{dk}
-f(\p)\frac{d\Gamma(E_{\p},k)}{dk}\right]
+\frac{1}{2}\nabla_{\p_T}^2[\delta\kappa_T(p) f(\p)] \, ,
\end{equation}
where the first term is calculated in the eikonal approximation, 
$\p+\k\simeq (p+k)\hat{\p}$, the last term is the radiative correction to 
the approximation, and $k<0$ corresponds to gluon absorption to satisfy 
detailed balance.  
The radiative correction to the transverse momentum diffusion coefficient is  
given by
\begin{equation}
\delta \kappa_T(p)
=\frac{g^2 C_F \kappa_{T}(p)}{2\pi}
\int\frac{dk}{k}[(1-x)^2+1][1+n_B(K)] 
\int\frac{d^2\k_T}{(2\pi)^2}\frac{k_T^2}{(\k_T^2+m^2x^2+m_g^2)^2} \, ,
\end{equation}
which is $\mathcal{O}(g^2)$ suppressed more than the leading-order 
coefficient but logarithmically enhanced \cite{Liou:2013qya}.

The heavy quark diffusion coefficient depending on momentum and temperature 
is important to determine the equilibration rate of heavy quarks in 
high-temperature QCD plasmas. 
The temperature dependence comes from running of the coupling constant: 
$\kappa_{L,T}\propto\alpha_s(E_{\p}T)\alpha_s(m_D^2)T^3$ \cite{Peigne:2008nd}, 
where $m_D$ is determined by 
$\ln\left(\frac{m_D^2}{\Lambda_{\rm QCD}^2}\right)
=\frac{N_c(1+N_f/6)}{11N_c-2N_f}\left(\frac{4\pi T}{m_D}\right)^2$ 
\cite{Peshier:2006ah}. 
In gluon-bremsstrahlung, the running coupling constant is determined at the 
scale, $Q^2=(\k_T^2+m^2x^2+m_g^2)/x$ \cite{Djordjevic:2013xoa}.
As the temperature decreases, the coupling becomes stronger and 
nonperturbative effects start to enter. 
The momentum diffusion coefficient of a heavy quark grows as the momentum 
increases. 
By employing the leading-log momentum dependence,
$\frac{\kappa_L(p)}{\kappa_L(p=0)}=\frac{3}{2}\left[\frac{E_{\p}^2}{p^2}-\frac{E_{\p}(E_{\p}^2-p^2)}{2p^3}\ln\frac{E_{\p}+p}{E_{\p}-p}\right]$
and 
$\frac{\kappa_T(p)}{\kappa_T(p=0)}=\frac{3}{2}\left[\frac{3}{2}-\frac{E_{\p}^2}{2p^2}+\frac{(E_{\p}^2-p^2)^2}{4E_{\p}p^3}\ln\frac{E_{\p}+p}{E_{\p}-p}\right]$ 
\cite{Moore:2004tg}, 
the static diffusion coefficient, $\kappa_{L,T}(p=0)$ which involves the 
nonperturbative effects, becomes the only transport parameter in 
Eq. (\ref{BoltzEq}). 
The momentum diffusion coefficient has been computed at next-to-leading order 
\cite{Caron-Huot:2007rwy}. 
Since the convergence of the perturbative expansion is poor for a realistic 
value of the strong coupling, it is desirable to determine it using 
nonperturbative approaches such as lattice QCD computation 
\cite{Banerjee:2011ra,Francis:2015daa,Brambilla:2020siz,Altenkort:2023eav} and 
phenomenological estimate \cite{ALICE:2021rxa,Xu:2017obm}. 
In this work, the momentum diffusion coefficient at $p=0$ and $T=T_c$ is 
constrained by lattice QCD data, while the momentum and temperature dependence 
are respectively determined by the leading-log result and running coupling.  
Through the lattice QCD data and the running coupling constant, 
nonperturbative effects can be absorbed at low momenta and temperatures near 
$T_c$.

\section{Nuclear modification factor and elliptic flow}
\label{result}

The heavy quark Boltzmann equation of Sec. \ref{formalism} has been solved 
for a Bjorken expansion to calculate the nuclear modification factor of 
heavy quarks \cite{Hong:2023cwl}. 
This formulation provides a consistent description of the collisional and 
radiative energy loss, which exhibit different momentum dependence in an 
intermediate-momentum regime. 
The suppression factor is primarily determined by the collisional and  
radiative energy loss in the low and high momentum limit, respectively. 
The significance of the radiative effect at intermediate momentum is found to 
be influenced by the momentum and temperature dependence of the transport 
coefficients. 
Although heavy quark approximations may be only marginally satisfied for 
charm quarks, the transport equation has been applied to both charm and 
bottom quarks in order to demonstrate the impact of mass. 
In this section, the same equation is solved in noncentral 
collisions to compute both the suppression factor and the anisotropic flow 
which are induced by the heavy quark energy loss.

The anisotropic flow of heavy quarks is generated by asymmetric energy loss 
in noncentral heavy-ion collisions. 
A relativistic hydrodynamic simulation provides the time 
evolution of the spatial distributions of temperature and flow 
velocity fields which are anisotropic in the transverse plane. 
As the temperature and flow velocity enter Eq. (\ref{BoltzEq}) through the 
heavy quark transport coefficients and the gluon emission rate, the Boltzmann 
equation becomes coupled with a hydrodynamic background. 
Initial conditions, especially the initial geometry and fluctuations of 
heavy-ion collisions, are a major source of significant uncertainties to 
determine QGP properties. 
Since this paper aims to qualitatively study how different mechanisms of 
energy loss influence the observables, a smooth initial distribution given by 
an optical Glauber model \cite{Miller:2007ri} is used. 
In a background medium generated by hydrodynamic simulations, 
Eq. (\ref{BoltzEq}) is solved with an initial condition where the momentum 
distribution is given by the differential cross section of heavy meson 
measured in pp collisions \cite{CMS:2017uoy,ALICE:2019nxm} and 
the spatial distribution is given by the same Glauber model,
\begin{equation}
f_{\x}(\x_T)=\int dz dz' \, 
\rho\big(\x_T+\frac{\b}{2},z\big)
\rho\big(\x_T-\frac{\b}{2},z'\big) \, ,
\end{equation}
where $\hat{z}$ is the beam axis and the impact parameter $\b$ is along 
the $\hat{x}$ direction. 
A Woods-Saxon form of density, $\rho(r)\propto 1/[1+\exp((r-R_0)/a)]$, 
is used with $R_0=6.62$ fm, $a=0.546$ fm for PbPb collisions.

\subsection{Expanding QGP}
\label{exp_qgp}

\begin{figure}
\includegraphics[width=0.45\textwidth]{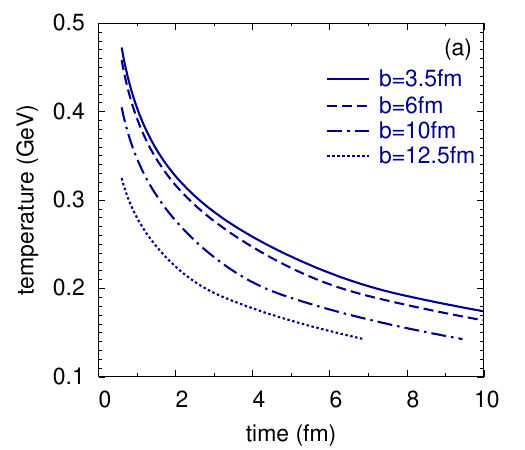}
\includegraphics[width=0.45\textwidth]{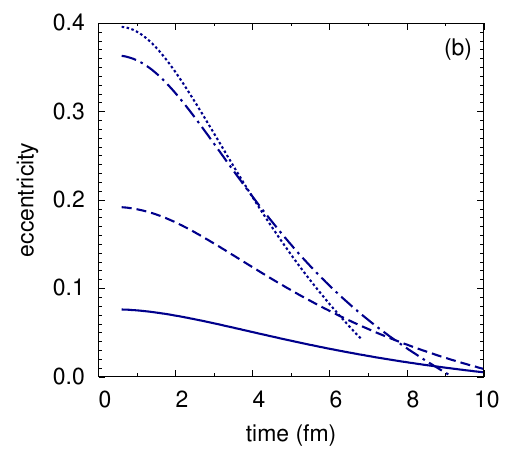}
\caption{
Evolution of (a) the temperature at the center of thermal media and (b) the 
spatial eccentricity in noncentral PbPb collisions with 
$b=3.5, \, 6, \, 10, \, \mbox{and} \, 12.5$ fm. 
}
\label{profile}
\end{figure}

In noncentral collisions with nonzero impact parameter, the elliptic 
deformation of thermal media can be quantified with the spatial 
eccentricity \cite{Kolb:2001qz}, 
\begin{equation}
\epsilon_2=\frac{\langle y^2-x^2\rangle}{\langle y^2+x^2\rangle} \, , 
\end{equation}
where the brackets denote averaging with the energy density as a weighting 
factor.
Because of the geometric deformation in the overlap region of two nuclei, 
the average velocity is larger in the $\hat{x}$ direction than $\hat{y}$. 
In the hydrodynamic picture, this spatial anisotropy is transferred to the 
momentum space through the pressure gradients \cite{Ollitrault:1992bk}. 
While the collective flow of light particles is effectively described by 
hydrodynamics, heavy quarks interacting with a background medium acquire 
various sources of azimuthal anisotropies through all stages of heavy-ion 
collisions. 
This work focuses on the energy-loss contribution to the heavy 
flavor flow in the QGP phase.

Figure \ref{profile} shows typical evolution of the QGP temperature  
at the center($x=0$, $y=0$) and the spatial eccentricity in noncentral PbPb 
collisions. 
They are simulated with (3+1)-dimensional hydrodynamics code MUSIC 
\cite{Schenke:2010nt,Schenke:2010rr,Paquet:2015lta} using the following 
hydrodynamic parameters\footnote{These parameters have 
been chosen to produce the maximum temperature similar to that in Ref. 
\cite{Alberico:2013bza}. 
Larger $e_0$ should generally be used to create higher-temperature 
plasmas at greater collision energies.}: the initial time, $t_0=0.6$ fm, 
the maximum energy density, $e_0=100$ GeV/fm$^3$, 
the ratio of shear viscosity to entropy density, $\eta/s=0.08$, and 
the equation of state from lattice QCD data \cite{Huovinen:2009yb}. 
In more central collisions with smaller impact parameters, the maximum 
temperature is higher and the initial spatial eccentricity is lower. 
While the temperature and the spatial eccentricity decrease with time,  
the momentum anisotropy is expected to increase. 
Since heavy quarks are not fully thermalized within the lifetime of QGP, they 
do not follow the bulk flow of a background medium. 
In response to an initial spatial deformation, heavy quarks acquire the 
momentum anisotropy by coupling to the collective motion of the medium and 
asymmetric energy loss. 
In the following subsections, the temperature and flow velocity fields, which 
depend on both space and time, are used to calculate the heavy quark flow and 
the suppression factor at midrapidity.

\subsection{Evolution of $R_{AA}$ and $v_2$}

\begin{figure}
\includegraphics[width=0.45\textwidth]{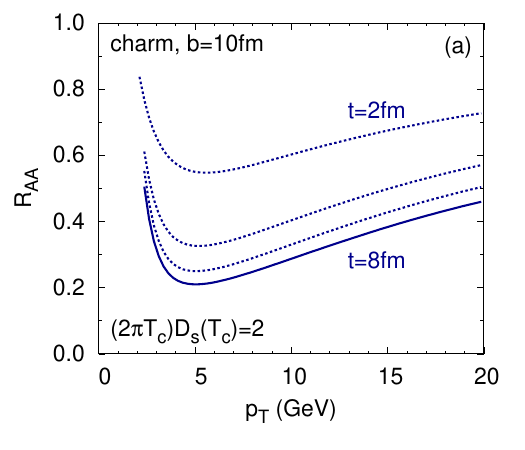}
\includegraphics[width=0.45\textwidth]{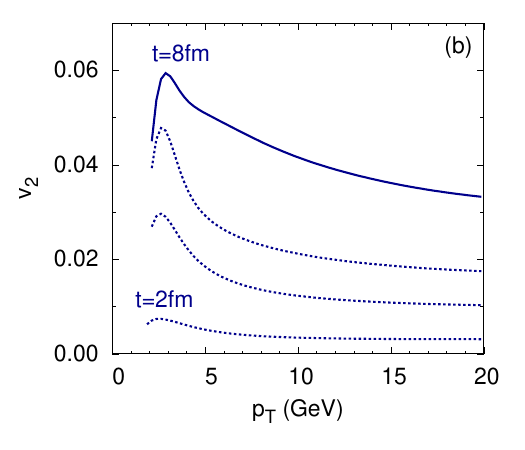}
\caption{
(a) The nuclear modification factor and (b) the elliptic flow of charm quarks  
at $t=2,\, 4, \, 6, \,\mbox{and}\, 8$ fm.  
}
\label{evol}
\end{figure}

At each position and time, heavy quarks experience the collisional and 
radiative energy loss which are controlled by the diffusion coefficient 
depending on temperature and flow velocity.
Once the temperature drops below $T_c\sim 157$ MeV \cite{HotQCD:2018pds}, 
heavy quarks are treated as free-streaming.  
After solving Eq. (\ref{BoltzEq}), the transverse momentum spectrum of heavy 
quarks is obtained by    
$\frac{dN}{d^2\p_T}=\int d^2\x_T \, f(t,\x_T,\p_T)$. 
The suppression factor is determined by the ratio of the quenched transverse 
momentum spectrum to the unquenched spectrum without the energy-loss effect.  
The collective flow is quantified in terms of Fourier components of the 
azimuthal angle distribution:
\begin{equation}
\frac{dN}{d^2\p_T}=
\frac{1}{2\pi}\frac{dN}{p_Tdp_T}\left[1+2\sum_n v_n(p_T)\cos n\phi\right] 
\, .
\end{equation}
Ignoring higher-order harmonics, 
$R_{AA}^{\rm in,out}(p_T)=R_{AA}(p_T)[1\pm 2v_2(p_T)]$, 
where the superscripts in $R_{AA}^{\rm in,out}$ denote in-plane($\phi=0$) and 
out-of-plane($\phi=\pi/2$).
The azimuthally averaged suppression factor is 
$R_{AA}(p_T)=[R_{AA}^{\rm in}(p_T)+R_{AA}^{\rm out}(p_T)]/2$, and 
the elliptic flow is estimated as \cite{Xu:2014ica,ALICE:2014qvj}
\begin{equation}
v_2(p_T)=\frac{1}{2}\frac{R_{AA}^{\rm in}(p_T)-R_{AA}^{\rm out}(p_T)}
{R_{AA}^{\rm in}(p_T)+R_{AA}^{\rm out}(p_T)} \, . 
\end{equation}
The nuclear modification factor of heavy quarks indicates the energy-loss 
effects in QGP, while the elliptic flow reflects how heavy quarks are 
distributed in the transverse plane and how they are correlated with the 
flowing thermal media.  
The anisotropic flow of heavy quark arises from the asymmetric energy loss, 
coupled with the collective motion of QGP: heavy quarks 
out-of-plane cross longer paths of hot media so they experience more energy 
loss than those in-plane. 
Since $R_{AA}^{\rm out}$ is smaller than $R_{AA}^{\rm in}$, heavy quarks 
obtain positive elliptic flow.

Figure \ref{evol} shows the evolution of the suppression factor and the 
anisotropic flow of charm quarks in noncentral collisions with $b=10$ fm. 
The charm-quark mass, $m=1.5$ GeV, and the spatial diffusion coefficient, 
$(2\pi T_c)D_s(T_c)=2$ where $D_s(T)=2T^2/\kappa_{L,T}(p=0,T)$, 
have been used.  
The $R_{AA}$ factor decreases with time and saturate at $t\approx 8$ fm 
when the plasma reaches the hadronization temperature. 
Meanwhile, $v_2$ continuously increases. 
The elliptic flow is affected by the different path-length as well as the 
anisotropic temperature and collective flow.   
In contrast to $R_{AA}$ which decreases mostly in the high-temperature plasmas 
at early times, $v_2$ buildup dominates at late times when the bulk flow is 
large and the heavy quark coupling with the medium is strong. 
Similar behavior in the time evolution has been observed previously 
\cite{Rapp:2008qc}. 
Although the interaction rate is high and the flow transfer is efficient in 
the early stages, it takes time for the flow to accumulate. 
Due to the development at later times, a significant contribution to $v_2$ 
is also expected at the hadronic phase. 
Furthermore, $v_2$ may be more sensitive than $R_{AA}$ to variations of $T_c$ 
and the uncertainties in the freezeout prescription.

\subsection{Collisional and radiative energy loss with $D_s$}
\label{ds_coeff}

The heavy quark formalism of Sec. \ref{formalism} consistently describes both 
diffusion and radiation processes through $D_s$ in an intermediate-momentum 
regime. 
The medium modification by gluon emission has been found to be 
distinguishable, exhibiting different momentum dependence compared to 
elastic scattering \cite{Hong:2023cwl}. 
In this subsection, the radiative effect on both $R_{AA}$ and $v_2$ is 
investigated in hydrodynamically expanding media.

\begin{figure}
\includegraphics[width=0.45\textwidth]{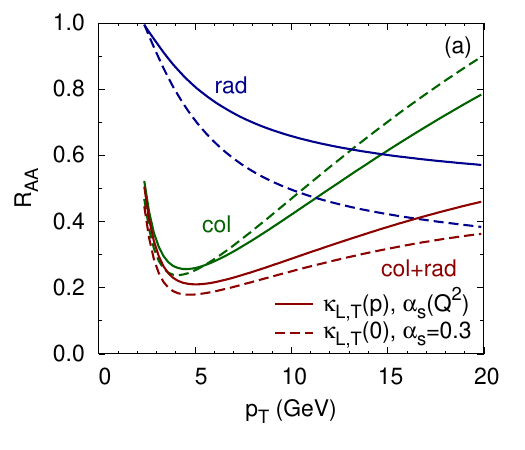}
\includegraphics[width=0.45\textwidth]{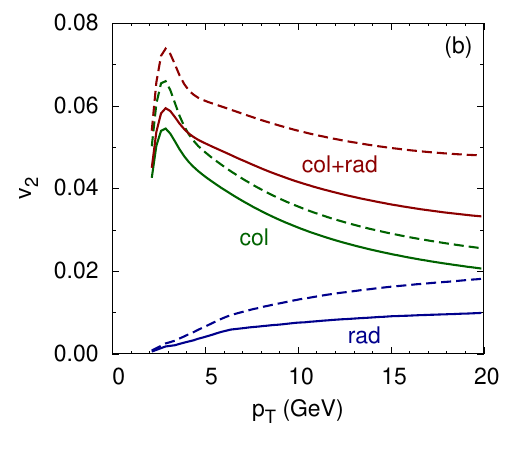}
\caption{
The collisional and radiative energy-loss contributions to (a) the nuclear 
modification factor and (b) the elliptic flow of charm quarks with 
$(2\pi T_c)D_s(T_c)=2$ in PbPb collisions with $b=10$ fm. 
The solid lines show the calculated results using the momentum-dependent 
$\kappa_{L,T}$ and running coupling $\alpha_s$, and the dashed lines show 
those with the constant diffusion coefficient and $\alpha_s=0.3$ for gluon 
emission.  
}
\label{trans_coeff}
\end{figure}

Figure \ref{trans_coeff} (a) shows the nuclear modification factor of charm 
quarks, determined by each energy-loss mechanism. 
The solid lines are the suppression factors calculated with the running 
coupling constant and the momentum-dependent diffusion coefficients, and 
the dashed lines are those with constant coupling and diffusion coefficient. 
Similar patterns of the medium modifications to Ref. \cite{Hong:2023cwl} have 
been observed with an adjustment to $D_s$. 
The $R_{AA}$ by the radiative energy loss starts at $R_{AA}=1$ for low momentum 
and consistently decreases as the momentum increases\footnote{
No gluon emission is expected when $p_T\lesssim m$.   
In the high-momentum limit, the gluon emission rate must 
be calculated in multiple soft scatterings, and the radiative energy loss 
will be reduced due to the LPM effect \cite{Landau:1953um,Migdal:1956tc}.}, 
while the $R_{AA}$ by the collisional energy loss decreases at 
low momentum but increases at intermediate momentum. 
As a result, the dominant energy loss shifts from collisional to radiative as 
the heavy quark momentum increases.  
The transition momentum is dependent on the transport coefficients and 
their dependence on momentum and temperature: 
since the radiative effect is more significant with $\alpha_s=0.3$ for gluon 
emission in the current numerical analysis, this transition occurs at 
higher momentum when $\kappa_{L,T}$ increases with momentum and $\alpha_s$ 
decreases with temperature, compared to when they are constant.

The collisional and radiative energy-loss contributions to the anisotropic 
flow are shown in Fig. \ref{trans_coeff} (b).
As discussed in the previous paragraph, the momentum dependence of two 
energy-loss mechanisms is qualitatively distinct.  
While the collisional energy loss is dominant at low $p_T$, the radiative 
effect increases with the heavy quark momentum. 
The $v_2$ of charm quarks peaks at $\sim 3$ GeV and decreases as the momentum 
increases: the radiative effect slows this decline. 
As in $R_{AA}$, the radiative energy loss with $\alpha_s=0.3$ is larger than 
that with running coupling.

In this paper, the heavy flavor observables have been calculated with 
the momentum-dependent diffusion coefficients and the running coupling 
constant, unless explicitly mentioned otherwise.

It should be mentioned that the valid range of the intermediate momentum, 
where gluon emission from a single scattering is applicable, is unclear. 
The momentum must be greater than the heavy quark mass but not so high that 
the LPM effect needs to be considered. 
Although gluon emission is more involved than photon emission, the LPM effect 
on the photon emission rate is less than $\sim 30\%$ \cite{Arnold:2001ms}. 
If this effect is included in the radiation term, the influence of the 
total energy loss decreases by $\sim 10\%$. 
However, the radiative energy loss still affects the $R_{AA}$ and $v_2$ 
in the way discussed above.

\subsection{Bulk flow effect}

To compute the $R_{AA}$ and $v_2$, the transverse momentum spectrum of heavy 
quarks is determined by heavy quark interactions in hydrodynamically expanding 
QGP. 
The interactions with a thermal medium can be characterized by the heavy 
quark diffusion coefficient, and its momentum and temperature dependence 
(through the running coupling constant) have been discussed in 
Sec. \ref{ds_coeff}. 
This subsection now focuses on the influence of an expanding medium.

\begin{figure}
\includegraphics[width=0.45\textwidth]{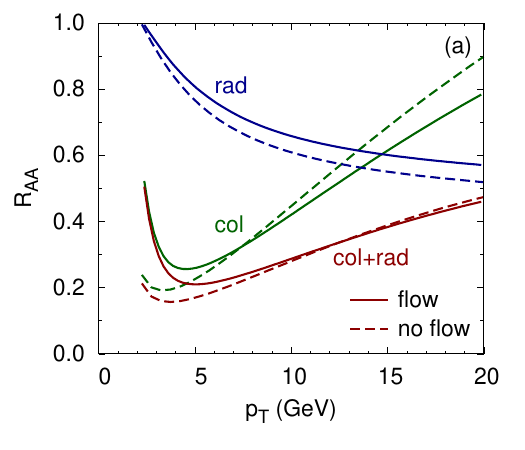}
\includegraphics[width=0.45\textwidth]{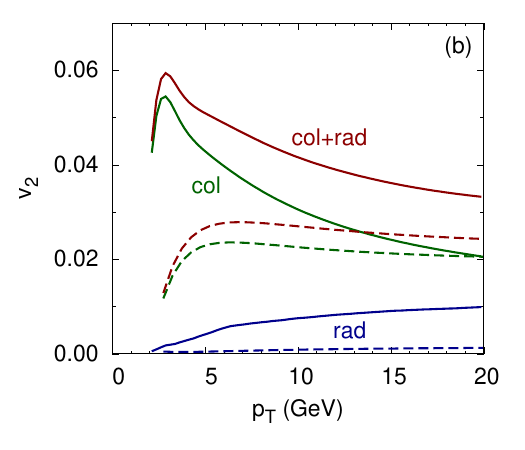}
\caption{
The bulk flow effect on (a) the suppression factor and (b) the anisotropic 
flow of charm quarks with $(2\pi T_c)D_s(T_c)=2$ for $b=10$ fm. 
The solid and dashed lines correspond to the results with and without the 
radial flow of an expanding medium, respectively.
}
\label{flow}
\end{figure}

Given $(t,\x_T)$, a hydrodynamic simulation provides the temperature and 
the flow velocity fields which are anisotropic in the transverse plane. 
For an expanding medium, the local rest frame (with primed coordinates) of the 
medium is different from the lab frame (with unprimed coordinates) where 
$\Delta \x_T=\frac{\p_T}{E_{\p}}\Delta t$ is satisfied. 
$\Delta t'$ and $\Delta \x_T'$ in the local rest frame are respectively given 
by $\Delta t'=\frac{E_{\p}'}{E_{\p}}\Delta t$ and 
$\Delta \x_T'=\frac{\p_T'}{E_{\p}}\Delta t$, where the energies and momenta in 
two frames are related through the Lorentz transformation: 
$E_{\p}'=\gamma(E_{\p}-\u_T\cdot\p_T)$ and 
$\p_T'=\p_T+\big(-\gamma E_{\p}+\frac{\gamma^2}{1+\gamma}\u_T\cdot\p_T\big)\u_T$ \cite{Baier:2006pt,Cao:2016gvr,Ke:2019jbh}.  
Transforming the heavy quark momentum from the lab frame to the local rest 
frame and using $\Delta t'$ and $\Delta\x_T'$ as given above, 
the heavy quark distribution is updated according to the Boltzmann equation 
in the local rest frame of the medium. 
The heavy quark diffusion coefficient which determines the energy loss 
depends on both the temperature and flow velocity of an expanding medium.
Thus, the transverse momentum spectrum in an expanding 
medium is influenced by bulk flow as well as by the asymmetric energy loss 
along different paths in an elliptically deformed medium. 
To investigate the contribution of bulk flow to the heavy flavor observables, 
the $R_{AA}$ and $v_2$ have been computed in the anisotropic 
temperature distribution of plasmas with and without the radial flow.

In Fig. \ref{flow}, the solid and dashed lines represent the results with and 
without the bulk flow, respectively. 
The bulk flow seems to shift the suppression factor to higher $p_T$, and the 
$R_{AA}$ becomes larger at low momentum. 
Since this effect is greater for in-plane than out-of-plane, 
the elliptic flow is also enhanced. 
The contribution of bulk flow is particularly significant to $v_2$ at low 
$p_T$ where $v_2$ exhibits a characteristic peak near $\sim 3$ GeV for 
charm quarks. 
On the other hand, the high-$p_T$ region is mainly determined by the 
asymmetric energy loss with the path-length difference due to the initial 
geometric deformation. 
The bulk flow effect on $R_{AA}$ and $v_2$ has been investigated in 
Refs. \cite{Cao:2016gvr,Nahrgang:2014vza}:
although the exact effect depends on the energy loss model, the shift of $R_{AA}$ to higher $p_T$ and 
the enhancement of $v_2$ at low $p_T$ are qualitatively consistent with the 
previous observations.

\subsection{Impact parameter and mass effect}

For the noncentral heavy-ion collisions in Sec. \ref{exp_qgp}, this subsection 
discusses the heavy flavor observables, focusing on the dependence on impact 
parameter and mass. 
The dependence on impact parameter is roughly related to the 
centrality dependence \cite{Broniowski:2001ei}. 
In realistic collisions with event-by-event fluctuations, heavy-ion 
collisions with the same impact parameter can produce different initial 
eccentricities. 
However, this work uses a smooth initial condition given by an optical 
Glauber model, so the eccentricity and the impact parameter are 
monotonically correlated.

\begin{figure}
\includegraphics[width=0.45\textwidth]{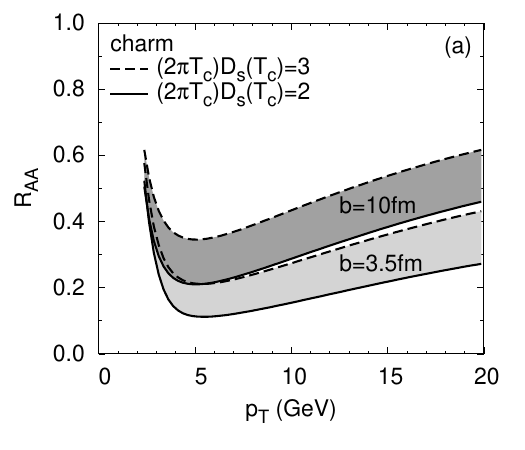}
\includegraphics[width=0.45\textwidth]{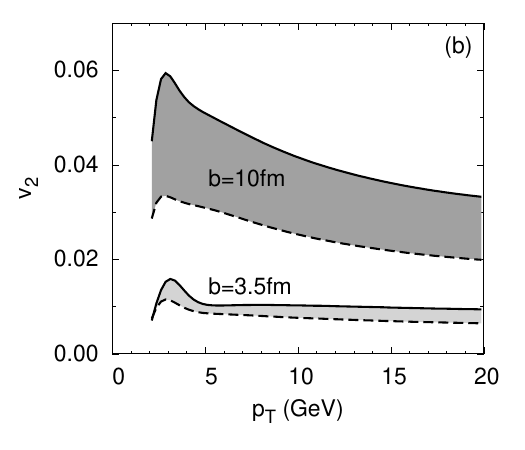}
\caption{ 
(a) The nuclear modification factor and (b) the elliptic flow of charm quarks 
with $(2\pi T_c)D_s(T_c)=2-3$ in noncentral PbPb collisions with 
$b=3.5, \, 10$ fm. 
}
\label{fig_c}
\end{figure}

Figure \ref{fig_c} shows the suppression factor and the elliptic flow of 
charm quarks with $(2\pi T_c)D_s(T_c)=2-3$ which are closely aligned with 
lattice QCD data 
\cite{Banerjee:2011ra,Francis:2015daa,Brambilla:2020siz,Altenkort:2023eav}. 
The heavy quark transport parameter determines the degree of heavy 
quark energy loss: as $D_s$ decreases, both the collisional and radiative 
energy loss increase. 
Thus, the nuclear modification factor becomes smaller, and the anisotropic 
flow increases for smaller $D_s$. 
For fixed $D_s$, the $R_{AA}$ factor at $b=3.5$ fm is smaller than at 
$b=10$ fm because the energy loss in more central collisions is larger due to 
bigger size and higher temperature of plasmas. 
On the other hand, the elliptic flow at $b=3.5$ fm is smaller than at 
$b=10$ fm because the initial eccentricity is greater in more peripheral 
collisions.

\begin{figure}
\includegraphics[width=0.45\textwidth]{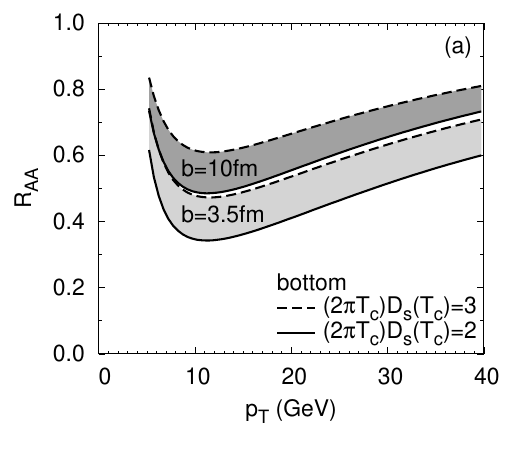}
\includegraphics[width=0.45\textwidth]{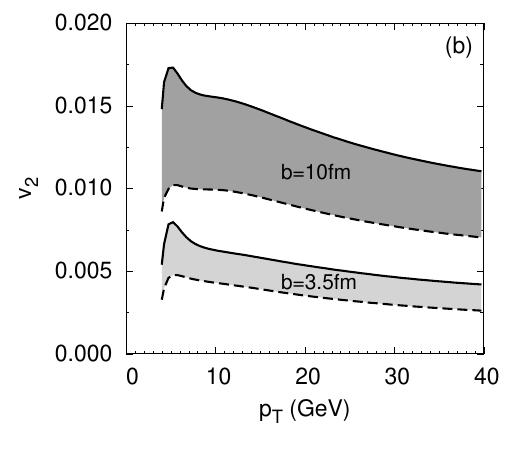}
\caption{ 
Same as Fig. \ref{fig_c} but for bottom quarks. 
}
\label{fig_b}
\end{figure}

The suppression factor and the anisotropic flow of bottom quarks are shown in 
Fig. \ref{fig_b}. 
Using the bottom-quark mass, $m=4.5$ GeV, and the same value for the diffusion 
coefficient, both the collisional and radiative energy loss of bottom quarks 
are smaller than those of charm quarks. 
Thus, bottom quarks are less suppressed and the momentum-space anisotropy is 
also weaker.

The $R_{AA}(p_T)$ and $v_2(p_T)$ of $D,\, B$ mesons have been measured  
\cite{CMS:2017uoy,ATLAS:2020yxw,ALICE:2021rxa,CMS:2024vip}, 
and bottom energy loss has been investigated through nonprompt $D$ meson 
\cite{ALICE:2022tji,ALICE:2023gjj} in PbPb collisions at LHC. 
Despite numerous efforts to develop the transport models of heavy quarks, it 
remains demanding to describe two quantities simultaneously. 
Since the observables of the heavy mesons are also influenced by factors 
other than the heavy quark interaction in QGP, the results of Figs. \ref{fig_c} 
and \ref{fig_b} should not be directly compared with the experimental data. 
However, their behavior, depending on the centrality(impact parameter), mass, 
and momentum, is similar to the measurements.    
In particular, the calculated suppression factors appear to be comparable with 
the data, whereas the prediction of the anisotropic flow is underestimated. 
Compared to $R_{AA}$, $v_2$ is more sensitive to the details of the transport 
models, such as the heavy quark transport coefficients and the 
centrality-dependent medium evolution with hydrodynamic flow 
\cite{Beraudo:2018tpr}. 
For example, given the same $R_{AA}(p_T)$, $2-3$ times larger $v_2(p_T)$ can 
be predicted by different temperature dependence of the heavy quark drag 
coefficient \cite{Das:2015ana}. 
Since the elliptic flow is primarily developed at the later stages of the QGP 
evolution, hadronization \cite{Rapp:2018qla,Zhao:2023nrz} and the freezeout 
prescription are also essential to determine $v_2$\footnote{The elliptic 
flow might arise from heavy quarks decoupling at different stages of QGP 
\cite{Beraudo:2017gxw}.}. 
Especially, coalescence considerably enhances the elliptic flow at low 
momentum \cite{He:2011qa,Beraudo:2022dpz}, transferring the collective flow 
from medium to heavy mesons. 
For a more quantitative analysis, realistic initial conditions, including 
fluctuations, should also be applied and the hydrodynamic parameters must be 
adjusted to describe the soft sector prior to heavy quark study 
\cite{Kurian:2020orp,Singh:2023smw}. 
This will be addressed in future work.

\begin{figure}
\includegraphics[width=0.45\textwidth]{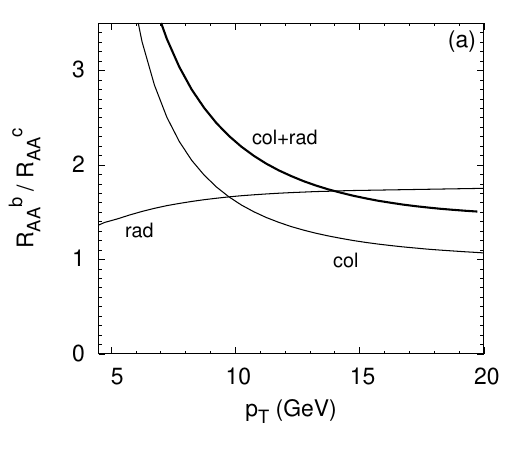}
\includegraphics[width=0.45\textwidth]{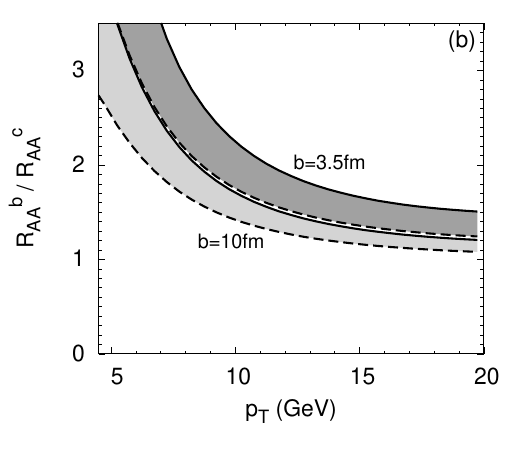}
\caption{
(a) The ratio of bottom to charm quark $R_{AA}$ at $b=3.5$ fm, determined by 
the different mechanisms of energy loss.  
(b) The ratios by the total energy loss at $b=3.5, \, 10$ fm, with 
$(2\pi T_c)D_s(T_c)=2$(solid lines) and $(2\pi T_c)D_s(T_c)=3$(dashed lines). 
}
\label{ratios}
\end{figure}

A comparison between the $R_{AA}$ factors of bottom and charm quarks 
is motivated by the measurements of the nonprompt to prompt 
$D^0$-meson $R_{AA}$ ratio \cite{ALICE:2022tji}. 
Fig. \ref{ratios} (a) shows the ratio of bottom to charm quark $R_{AA}$, which 
implies the mass-dependent energy loss. 
Because the energy loss of charm quarks are larger than that of bottom quarks, 
$R_{AA}^b/R_{AA}^c$ is bigger than unity. 
In particular, the radiative effect becomes significant to determine the ratio 
as the momentum increases. 
The ratio given by the total energy loss is notably enhanced at low $p_T$, 
then decreases as $p_T$ increases, approaching unity: an analogous result has 
been obtained in Ref. \cite{Li:2020umn} excluding coalescence 
which is effective at low momentum.  
The average ratio of the nonprompt to prompt $D^0$-meson $R_{AA}$ is 
$1.70\pm0.18$ \cite{ALICE:2022tji} which is comparable with the calculated 
$R_{AA}^b/R_{AA}^c$ near $p_T\sim20$ GeV (where the coalescence effect is 
negligible). 
In more peripheral collisions where the energy-loss difference between bottom 
and charm quarks is smaller, the ratio is reduced, as shown in 
Fig. \ref{ratios} (b).

\begin{figure}
\includegraphics[width=0.45\textwidth]{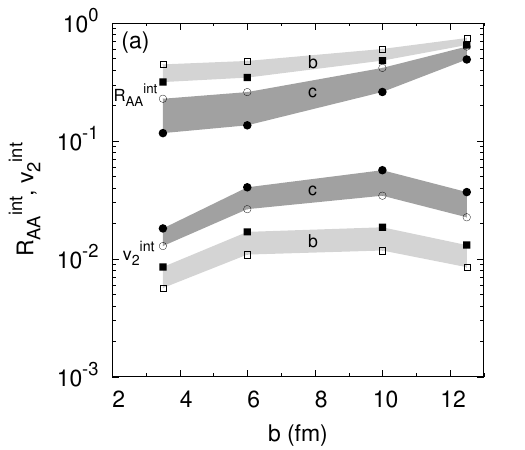}
\includegraphics[width=0.45\textwidth]{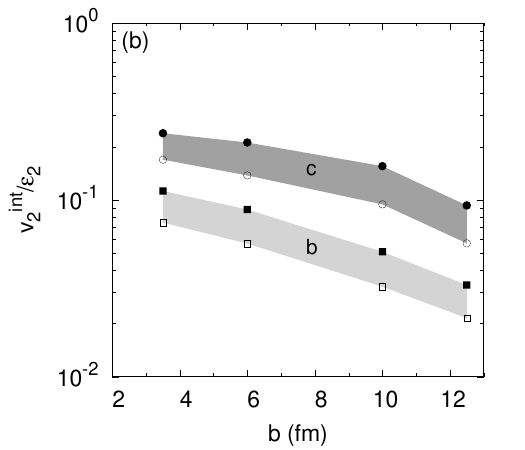}
\caption{
(a) The $p_T$-integrated $R_{AA}$ and $v_2$ as functions of impact parameter. 
(b) The ratio of the $p_T$-integrated $v_2$ to the initial spatial 
eccentricity. 
The constant transport coefficients 
$(2\pi T)D_s(T)=2$ (closed symbols), $(2\pi T)D_s(T)=3$(open symbols), 
and $\alpha_s=0.3$ for gluon emission have been used. 
}
\label{corr_scale}
\end{figure}

To closely investigate the correlation between $R_{AA}$ and $v_2$, and their 
dependence on $D_s$, mass, and impact parameter, 
the $p_T$-integrated elliptic flow is defined as follows: 
\begin{equation}
v_2^{\rm int}(b)=
\frac{\int dp_T \, p_T v_2(p_T,b)\frac{1}{2\pi}\frac{dN}{p_T dp_T}}
{\int dp_T \, p_T \frac{1}{2\pi}\frac{dN}{p_T dp_T}} \, , 
\end{equation}
where $4<p_T<20$ GeV, and similarly for $R_{AA}^{\rm int}(b)$. 
Fig. \ref{corr_scale} (a) shows the integrated $R_{AA}$ and $v_2$ of 
charm and bottom quarks in four noncentral heavy-ion collisions. 
While $R_{AA}$ and $v_2$ as functions of $p_T$ depend on the 
specifics of the heavy quark interaction in expanding QGP, the 
$p_T$-integrated observables generally exhibit the following 
centrality-dependent behavior. 
The integrated $R_{AA}$ increases with $b$ as the energy loss 
is larger in more central collisions where QGP have bigger 
size and higher temperature(longer lifetime)\footnote{See Fig. 
\ref{profile} (a). The initial overlap area of two nuclei at $b=3.5$ fm is 
about $3-4$ times as large as that at $b=12.5$ fm.}. 
On the other hand, the integrated $v_2$ grows with $b$ up to $b\sim 10$ fm 
after which it decreases for more peripheral collisions. 
The elliptic flow of heavy quarks is influenced by two competing effects, 
the geometric deformation and energy loss \cite{Xing:2024qcr}. 
As $b$ increases, the geometric asymmetry becomes bigger while the energy 
loss effect becomes smaller. 
A similar centrality-dependent behavior has been shown in the 
$p_T$-dependent $R_{AA}$ and $v_2$ of Ref. \cite{Xing:2024qcr}. 
The $D_s$ and mass dependence of the integrated $R_{AA}$ and $v_2$ remain the 
same as discussed in the $p_T$-dependent observables.

The heavy quark interaction and the coupling to QGP are weaker in more 
peripheral collisions \cite{Nahrgang:2014vza}. 
As the impact parameter increases, the efficiency of transferring an initial 
geometric deformation to the momentum-space anisotropy decreases.
Fig. \ref{corr_scale} (b) shows the ratio of the $p_T$-integrated $v_2$ to 
the initial spatial eccentricity. 
The $v_2^{\rm int}/\epsilon_2$ decreases as the impact parameter increases, 
and the ratio of bottom quarks is smaller than that of charm quarks.

\section{summary}
\label{summary}

To demonstrate the transition between the collisional and radiative energy 
loss, heavy quark suppression and anisotropic flow have been computed in an 
intermediate-momentum regime where the finite mass is nonnegligible. 
For diffusion and semi-collinear gluon-bremsstrahlung, the qualitative 
features of the two energy-loss mechanisms have been consistently investigated  
by using the heavy quark diffusion coefficient of lattice QCD data.
The $p_T$-dependent $R_{AA}$ and $v_2$ induced by the radiative energy loss 
are found to be distinguishable from those caused by the collisional energy 
loss. 
By slowing the increase of $R_{AA}(p_T)$ and the decreases of $v_2(p_T)$, the 
radiative effect makes the observables more independent of $p_T$ in the 
intermediate-momentum region, compared to those entirely determined by the 
collisional effect.

The numerical analysis indicates that a significant part of the heavy meson 
suppression results from the energy loss of heavy quarks particularly in the 
early stages of QGP. 
In contrast, the elliptic flow is influenced by the interplay between 
the energy loss and the spatial eccentricity in noncentral heavy-ion 
collisions, with its growth being considerable at late times.    
The contribution of bulk flow is found to be important at low momentum, 
especially to $v_2$. 
The energy-loss difference between charm and bottom, as well as the 
centrality-dependence of the observables, has been discussed.

Reference \cite{Hong:2023cwl} has presented a transport formulation for 
heavy quark diffusion and radiation with a single transport parameter, and 
this work conducts the numerical analysis in hydrodynamically expanding media. 
They provide a useful approach to exploring the intermediate-momentum regime, 
where the dominant energy loss shifts from elastic scattering to 
gluon-bremsstrahlung. 
The distinct medium modifications help clarify the relative importance of the 
two mechanisms and their influence on the heavy flavor observables. 
The observables are primarily determined by the collisional energy loss at 
low momentum, while the influence of the radiative effect increases as 
the heavy quark momentum grows. 
Although several transport models of heavy quarks are available to describe 
experimental data, this approach may offer a new perspective on the 
emergence of the radiative effect in the heavy quark energy loss.

\section*{Acknowledgments}

I would like to thank Sangyong Jeon and Su Houng Lee for useful comments. 
This work was supported by the National Research Foundation of Korea(NRF) 
grant funded by the Korea government(MSIT) (RS-2024-00342514) and 
Basic Science Research Program funded by the Ministry of Education 
(No. 2021R1I1A1A01054927).

\end{document}